\documentclass[a4paper]{jpconf}
\usepackage{graphicx}
\begin{document} 
\title{Three form potential in (special) minimal supergravity superspace and supermembrane supercurrent}

\author{Igor A. Bandos $^\dagger$$^\ast$ and
 Carlos Meliveo
 $^\ast$}
  \address{$^\dagger$IKERBASQUE, the Basque Foundation for Science, Bilbao,
Spain \\   $^\ast$Department of Theoretical Physics, University of the Basque Country,   P.O. Box 644, 48080 Bilbao,
Spain }
\date{V1: July 16, 2011. V2: October 24, 2011. Printed \today }

\ead{igor\_bandos@ehu.es, bandos@ific.uv.es, carlos\_meliveo@ehu.es. }

\begin{abstract}This contribution begins the study of the complete superfield Lagrangian description of the interacting system of D=4 ${\cal N}=1$ supergravity (SUGRA) and supermembrane.
Firstly, we review a 'three form supergravity' by Ovrut and Waldram, which we prefer to call 'special minimal supergravity'. This off--shell formulation of simple SUGRA is appropriate for our purposes as the supermembrane action contains the so--called Wess-Zumino term given by the integral over a three form potential in superspace, $C_3$. We describe this formulation in the frame of Wess--Zumino superfield approach, showing how the basic variations of minimal SUGRA are restricted by the conditions of the existence of a three-form potential $C_3$ in its superspace. In this language the effect of dynamical generation of cosmological constant, known to be characteristic for this formulation of  SUGRA, appears in its superfield form, first described by Ogievetsky and Sokatchev in their formulation of SUGRA as a theory of  axial vector superfield. Secondly, we vary the supermembrane action with respect to the special minimal SUGRA superfields (basic variations) and  obtain the supercurrent superfields as well as the supergravity superfield equations with the supermembrane contributions.\end{abstract}

\date{V1: July 16, 2011. V2: October 27, 2011. Printed \today }



\section{Introduction.}

Recently the increasing interest in the superfield formalism in flat and curved $D=4$, ${\cal N}=1$ superspaces can be witnessed  \cite{Seiberg+ZK=2009,Seiberg+2009,Goldstino09-,Seiberg+2010,Kuzenko:2010,Seiberg+2011,Seiberg+=2011,Bagger:2011na}. In particular, re-addressing the old problems of the superfield description of the Goldstino \cite{GoldstinoSuperfield}, which is the Goldstone fermion related to the spontaneous supersymmetry breaking \cite{Volkov73}, and of the supercurrent superfield \cite{FerraraZumino74}, allowed the authors of \cite{Seiberg+2009}  to develop a  systematic approach to describe low-energy couplings of Goldstinos the results of  which have been already applied including to the development of a new 'minimal inflation' scenario \cite{AlvarezGaume:2010rt}. The use of auxiliary fields of supergravity multiplets as ''background fields'' allowed \cite{Seiberg+=2011} to simplify essentially the derivation of curved space versions of rigid supersymmetric theories.  A relation of different supercurrents with brane currents corresponding to the topological charges in supersymmetry algebra \cite{Jose+Paul=PRL89} was discussed in recent \cite{Seiberg+2011}.

Actually, the supercurrent superfields produced by the $D=4$, ${\cal N}=1$ superparticle, superstring and supermembrane can be calculated directly by varying their actions   \cite{BS81,G+S84,BST87,AGIT=88} in an off-shell supergravity background with respect to the pre-potentials of superfield supergravity \cite{OS78,Siegel:1978nn,Siegel79,OS80} (see also \cite{1001,BK,SiBook}) or, equivalently, with respect to the constrained supervielbein using the technique of \cite{WZ78}\footnote{The Wess--Zumino approach to minimal supergravity \cite{WZ77,GWZ78}, \cite{WZ78,WZ79} is the subject of the excellent book \cite{BW} which, however, does not address the problem of varying the superfield supergravity action. Additional technical details on the superfield variations can be found in \cite{BdAIL03,IB+JI=03}.}.

Such a problem can be considered as a part of derivation of the superfield equations of the interacting system of $D=4$ ${\cal N}=1$   super--$p$--brane and dynamical supergravity  which have been addressed in  \cite{BdAIL03} and  \cite{IB+JI=03} for the case of $p=0$ and $p=1$, respectively. Thus the explicit expression for the supercurrent superfield of the massless superparticle and of the $D=4$ ${\cal N}=1$ Green--Schwarz superstring can be extracted from \cite{BdAIL03} and \cite{IB+JI=03}.

In this contribution  we obtain the $D=4$, ${\cal N}=1$ supermembrane current superfields\footnote{See \cite{IB+CM=2010} for the Lagrangian description of the interacting system of $D=4$, ${\cal N}=1$ supermembrane and scalar multiplet, which can be considered as a preparatory step of our present study. }.
To this end we needed to vary the supermembrane action with respect to the supergravity superfields, and,
firstly, to find an appropriate off--shell formulation of supergravity.
Thus it should not be surprising that a significant part of this contribution is devoted to
reviewing the issues of $D=4$, ${\cal N}=1$ superfield supergravity. Namely, we will describe, in the language of the Wess--Zumino superfield approach, the 'three form supergravity' by B. Ovrut and D. Waldram \cite{Ovrut:1997ur}, which we prefer to call 'special minimal supergravity'\footnote{In \cite{Kuzenko+05} the 'three-form supergravity' of B. Ovrut and D. Waldram was re-discovered in the frame of prepotential approach, where it clearly appears as a 'special' case of minimal supergravity formulation with chiral compensator constructed from the real rather than complex prepotential.}.


The supermembrane action \cite{BST87,AGIT=88} contains the Wess--Zumino term which is given by integral over the supermembrane worldvolume $W^3$ of a  3--form gauge potential $C_3={1\over 3!}dZ^M\wedge dZ^N\wedge dZ^L C_{LNM}(Z)$ defined on tangent superspace. The three form matter multiplet in the superspace of minimal supergravity was studied in  \cite{Binetruy:1996xw}. When the supermembrane interaction with supergravity is considered, $C_3$ should be constructed from the supergravity superfields.
The minimal $D=4$, ${\cal N}=1$ supergravity allows for the existence on its curved superspace of a closed 4--form which may be associated with the field strength of such a three form potential, $H_4=dC_3$ \cite{Ovrut:1997ur}.

However the attempt to re-construct the potential $C_3$ starting from the expression for $H_4$ in terms of supergravity superfields results in a modification of the auxiliary field sector of supergravity \cite{Ovrut:1997ur}. In the language of  prepotential  formulation of supergravity \cite{OS78,Siegel:1978nn,Siegel79} this can be described by saying that the compensator superfield of minimal supergravity (\cite{Siegel:1978nn}) becomes {\it special chiral superfield} which is constructed from the real pre-potential rather than from the complex one \footnote{Notice also that the coupling of supermembrane to scalar supermultiplet requires this to be described by special scalar superfield \cite{IB+CM=2010}. The appearance of special chiral superfield in the description of 3-form in flat superspace is noticed in \cite{Gates:1980ay}, \cite{Gates:1980az} and \cite{Binetruy:1996xw}.}. In the language of component tensor calculus \cite{TenCal}, the modification of the auxiliary field sector can be described as substituting a divergence of an auxiliary vector field, $\partial_\mu k^\mu$, (or, equivalently, the field strength of a 3-form field $\epsilon^{\mu\nu\rho\sigma}\partial_\mu B_{\nu\rho\sigma}$ \cite{Ovrut:1997ur}) for one of the auxiliary scalar  fields  of minimal supergravity. Although on the first glance  such a modification does not look essential,  it changes  the supergravity superfield equations by introducing an arbitrary real constant $c$ in the right hand side of the scalar superfield equation: $R=0\mapsto R=c$. This, in its turn, results in that the Einstein equation obtains a constant right hand side proportional to the square of  this integration constant,  $\propto c^2$, so that its ground state solution becomes $AdS_4$ (with a dynamically generated radius) rather than Minkowski space. Such a  {\it dynamical generation of cosmological constant} in superfield supergravity was observed for the first time by Ogievetski and Sokatchev \cite{OS80} in their formulation of supergravity as a theory of axial vector superfield \cite{OS78}. However, in more general perspective, the history of the  effect of dynamical generation of cosmological constant is much longer: it also occurs in the frame of so--called 'unimodular gravity', first considered by Einstein as early as in 1919 \footnote{See \cite{UnimGrav} for review and original references. The authors thank Dario Francia and Victor Revelles for giving them know about the existence of this model.}.
The fact that the supermembrane coupling to dynamical supergravity requires this to have an auxiliary field sector which results in dynamical generation of cosmological constant \cite{Ovrut:1997ur} looks natural if we remember that the bosonic body of supermembrane is a domain wall which may separate two spacetime regions with different values of cosmological constant.

The review of special minimal supergravity in Sections 2-4 provides the basis for our study of the superfield Lagrangian description of the supergravity---supermembrane interacting system.
In sec. 5, varying the supermembrane actions with respect to superfields of special minimal supergravity we find the explicit form of the supermembrane current superfields.  This and the superfield supergravity equations with the supermembrane source terms are the main results of the present contribution. The detailed study of the structure of these  superfield equations  will be the subject of forthcoming paper. Our conclusions and discussion can be found in sec. 5.

\section{Three form potential in curved $D=4$, ${\cal N}=1$ superspace and special minimal supergravity.}

\medskip 

\subsection{Supermembrane action and three form potential in curved superspace.}

\medskip 

We denote local coordinates of the curved $D=4$ ${\cal N}=1$ superspace $\Sigma^{(4|4)}$  by
$\{ {Z}^M\} \equiv   \{ x^\mu, \theta^{\underline{\breve{\alpha}}} \}$ ($\mu=0,1,2,3$, $\underline{\breve{\alpha}}=1,2,3,4$), and the bosonic and fermionic supervielbein one forms of  $\Sigma^{(4|4)}$ by
\begin{eqnarray}\label{4Ea}  E^a&=& dZ^M E_M^a(Z)\; , \quad
E^{{\alpha}}= dZ^M E_M^{{\alpha}}(Z)\; , \quad
 \bar{E}{}^{\dot\alpha}= dZ^M \bar{E}_M{}^{\dot\alpha}(Z)\; . \quad
\end{eqnarray}
Here $a=0,1,2,3$, $\alpha =1,2$, $\dot{\alpha}=1,2$. Sometimes it is convenient to collect the supervielbein one forms in $E^{A} = ( E^a, E^{{\alpha}}, \bar{E}_{{\dot\alpha}})$.

As it is well known, the supermembrane action \cite{BST87,AGIT=88} is given by the sum of the Dirac--Nambu--Goto and the Wess--Zumino term,
  \begin{eqnarray}\label{Sp=2:=}
  S_{p=2}= {1\over 2}\int d^3 \xi \sqrt{g} - \int\limits_{W^3} \hat{C}_3 = \; \qquad \nonumber \\ = -{1\over 6} \int\limits_{W^3} *\hat{E}_a\wedge \hat{E}{}^a - \int\limits_{W^3} \hat{C}_3\; . \qquad
\end{eqnarray}
The former is given by the volume of $W^3$ defined as  integral of the  determinant of the induced metric, $g=det(g_{mn})$,
\begin{eqnarray}\label{g=EE}
g_{mn}= \hat{E}_m^{a}\eta_{ab} \hat{E}_n^{b}\; , \qquad
\hat{E}_m^{a}:= \partial_m \hat{Z}^M(\xi) E_M^a(\hat{Z})\; .  \qquad
\end{eqnarray}
Here $\xi^m= (\tau,\sigma^1, \sigma^2)$ are local coordinates on $W^3$ and $\hat{Z}^M(\xi)$ are coordinates functions which determine the embedding of $W^3$ as a surface in target superspace $\Sigma^{(4|4)}$,
\begin{eqnarray}\label{W3inS44}
W^3\; \subset  \Sigma^{(4|4)}\; : \qquad Z^M= \hat{Z}{}^{{ {M}}}(\xi)= (\hat{x}{}^{\mu}(\xi)\, ,
\hat{\theta}^{\breve{\alpha}}(\xi))\; . \;
\end{eqnarray}
In the second line of Eq. (\ref{Sp=2:=}) the Dirac--Nambu--Goto term is written  as an integral of the wedge product of the pull--back of the $\Sigma^{(4|4)}$ bosonic supervielbein form $E^a$ to $W^3$,
\begin{eqnarray}\label{hEa=dxiE}
\hat{E}{}^a= d\xi^m \hat{E}_m^{a} = d \hat{Z}^M(\xi) E_M^a(\hat{Z})\; ,  \qquad
\end{eqnarray}
and of its Hodge dual two form defined with the use of the induced metric (\ref{g=EE}) and its inverse $g^{mn}$,
\begin{eqnarray}\label{*Ea:=}
*\hat{E}^a:= {1\over 2}d\xi^m\wedge d\xi^n\sqrt{g}\epsilon_{mnk}g^{kl}\hat{E}_l^a \; . \qquad \end{eqnarray}

The second, Wess--Zumino term of the supermembrane action (\ref{Sp=2:=}) describes the  supermembrane coupling to a three--from gauge potential defined on $\Sigma^{(4|4)}$. Thus, to write a supermembrane action, one has to construct the three--form gauge potential $C_3$ in the target superspace $\Sigma^{(4|4)}$. In the case of flat target superspace the exterior derivative of such a three form $C_3^0$ will define a real Chevalley-Eilenberg (CE) 4-cocycle $H^0_4$ \cite{JdA+PKT89}, {\it i.e.} a closed ($dH^0_4=0$) and supersymmetric invariant four form. Although $H^0_4$ is the complete derivative of a three--form, $H^0_4=dC^0_3$ (and thus is trivial in de Rahm cohomology), as far as $C_3^0$ is not invariant, but rather changes on a complete derivative under the supersymmetry transformations \cite{AGIT=88} (see also \cite{IB+CM=2010}), $H^0_4$ is a nontrivial 4-cocycle in the Chevalley--Eilenberg cohomology \cite{JdA+PKT89,Jose+Paul=PRL89,AI95}.

To stress the importance of closed four form $H^0_4=dC^0_3$, let us notice that, sometimes one writes the  Wess-Zumino term as an integral of this four form $H^0_4$ over a 4 dimensional manifold $\tilde{W}^4$ the boundary of which is the supermembrane worldvolume $W^3$, $\int_{W^3}\hat{C}_3= \int_{\tilde{W}{}^4} \tilde{H}^0_4$. (Certainly, to make such a statement precise, one should redefine the system a bit,  but we will not be needing this below).

\subsection{Supermembrane  and closed 4--form in the minimal supergravity  superspace. }

\medskip 

Thus, to construct the  supermembrane action in a supergravity background one has to construct first the closed form $H_4$ ($dH_4=0$) in curved supergravity superspace.
The superspace of minimal supergravity allows for existence of two closed 4-forms
\begin{eqnarray} \label{H4L}  & H_{4L} =  - {i\over 4} E^b\wedge E^a \wedge E^\alpha \wedge E^\beta \sigma_{ab\; \alpha\beta} -   {1\over 128} E^{d} \wedge E^c \wedge E^b \wedge E^a \epsilon_{abcd} R \, , \qquad
 dH_{4L}=0 \, ,
\end{eqnarray}
and its complex conjugate $H_{4R}=(H_{4L})^*$ \footnote{The real form $H_L+H_R$ was described in \cite{Ovrut:1997ur} based on the previous study of \cite{Binetruy:1996xw}.  We are thankful to Daneil Waldram for bringing these important references to our attention.}.

To show that the 4--form (\ref{H4L}) is closed, $dH_{4L}=0$, one has to use the superspace constraints of minimal supergravity, which can be collected in the expressions for the bosonic and fermionic torsion 2--forms,
\begin{eqnarray}\label{4WTa=}T^a &:=& {\cal D}E^a
=- 2i\sigma^a_{\alpha\dot{\alpha}} E^\alpha \wedge \bar{E}^{\dot{\alpha}} - {1\over 8} E^b \wedge E^c
\varepsilon^a{}_{bcd} G^d \; ,\;   \\ \label{4WTal=} T^{\alpha} &:=&{\cal D}E^\alpha  = {i\over 8} E^c \wedge E^{\beta}
(\sigma_c\tilde{\sigma}_d)_{\beta} {}^{\alpha} G^d   -{i\over 8} E^c
\wedge \bar{E}^{\dot{\beta}} \epsilon^{\alpha\beta}\sigma_{c\beta\dot{\beta}}R +
 {1\over 2} E^c \wedge E^b \; T_{bc}{}^{\alpha}\; ,\;
\end{eqnarray}
as well as some of the properties of main superfields entering (\ref{4WTa=}), (\ref{4WTal=})
\footnote{Our notation coincides with that of \cite{BdAIL03} and \cite{IB+CM=2010} (and is quite close to that of \cite{BW}, although our metric is mostly minus, $\eta^{ab}=diag (1,-1,-1,-1)$). The covariant derivative ${\cal D}=E^a{\cal D}_a+ E^\alpha {\cal D}_\alpha + \bar{E}{}^{\dot{\alpha}}\bar{{\cal D}}_{\dot{\alpha}}$ involves the spin connection $\omega^{ab}=-\omega^{ba}=dZ^M\omega_M^{ab}(Z)$, {\it e.g.} ${\cal D}E^a= dE^a-E^b\wedge \omega_b{}^a$ and  ${\cal D}E^\alpha= dE^\alpha-E^\beta\wedge \omega_\beta{}^\alpha$ with $\omega_\beta{}^\alpha={1\over 2}\omega^{ab}\sigma_{ab}{}_\beta{}^\alpha$; the exterior derivative acts from the right, so that, {\it e.g.}
 ${\cal D}(E^b\wedge E^\alpha
)= E^b\wedge {\cal D}E^\alpha - {\cal D}E^b\wedge E^\alpha$.} explicitly,
\begin{eqnarray} \label{chR} & {\cal D}_\alpha \bar{R}=0\;
, \qquad \bar{{\cal D}}_{\dot{\alpha}} {R}=0\; ,
\\
\label{DG=DR} & \bar{{\cal
D}}^{\dot{\alpha}}G_{{\alpha}\dot{\alpha}}= - {\cal D}_{\alpha} R \; , \qquad {{\cal
D}}^{{\alpha}}G_{{\alpha}\dot{\alpha}}= - \bar{{\cal D}}_{\dot{\alpha}} \bar{R} \; .
 \qquad \end{eqnarray}
Notice that one more representative of the set of main superfields enter Eq.  (\ref{4WTal=}) implicitly, through the decomposition of the superfield generalization of the gravitino field strength $T_{bc}{}^{\alpha}(Z)$,
\begin{eqnarray}\label{Tabg} T_{{\alpha}\dot{\alpha}\; \beta \dot{\beta }\; {\gamma}} &
\equiv   \sigma^a_{{\alpha}\dot{\alpha}} \sigma^b_{\beta \dot{\beta }} \epsilon_{{\gamma}{\delta}}
T_{ab}{}^{{\delta}}=  -{1\over 8}  \epsilon_{{\alpha}{\beta}} {\bar{{\cal D}}}_{(\dot{\alpha}|} G_{\gamma
|\dot{\beta})} - {1\over 8} \epsilon_{\dot{\alpha}\dot{\beta}}[W_{\alpha \beta\gamma} -
2\epsilon_{\gamma (\alpha}{\cal D}_{\beta)} R]
 \; . \end{eqnarray}
It is symmetric in spinorial indices, $W_{\alpha \beta\gamma}=W_{(\alpha \beta\gamma )}$, and obeys
 \begin{eqnarray}
 \label{chW} & \bar{{\cal D}}_{\dot{\alpha}} W^{\alpha\beta\gamma}= 0\; , \qquad
{{\cal D}}_{{\alpha}}\bar{W}^{\dot{\alpha}\dot{\beta}\dot{\gamma}}= 0\;,
 \\ \label{DW=DG} & {{\cal D}}_{{\gamma}}W^{{\alpha}{\beta}{\gamma}}= \bar{{\cal D}}_{\dot{\gamma}} {{\cal
D}}^{({\alpha}}G^{{\beta})\dot{\gamma}} \; . \qquad
\end{eqnarray}
Furthermore, it contains (spin--tensor representation of the) Weyl tensor, $C_{\alpha\beta{\gamma}{\delta}}=
C_{(\alpha\beta{\gamma}{\delta})}$ among its components. To be more precise, the superfield generalization of the Weyl tensor appears as one of the irreducible components of the  nonvanishing spinor derivative of
${W}_{{\alpha}{\beta}{\gamma}}$,
\begin{eqnarray} \label{Weyl} & C_{\alpha\beta{\gamma}{\delta}}:=
r_{(\alpha\beta{\gamma}{\delta})} = -  {1\over 16} {{\cal D}}_{({\alpha}} W_{\beta{\gamma}{\delta})} \; .
\end{eqnarray}

The real part of the complex 4-form $H_{4L}$ in (\ref{H4L}),
\begin{eqnarray} \label{H4=HL+HR}  H_4&:=&dC_3= H_{4L}+H_{4R}\;  ,   \qquad
\end{eqnarray}
is also closed  and provides the natural candidate for the four form associated to the Wess--Zumino term of the supermembrane $\int\limits_{W^3}{C}_3$ in (\ref{Sp=2:=}) \cite{Ovrut:1997ur}. To obtain the supermembrane equations of motion in a supergravity background and to check for its gauge symmetries it is sufficient to know  the form of $H_4$ in terms of supergravity superfields, Eq. (\ref{H4=HL+HR}), (\ref{H4L}).

However, to calculate the supermembrane current(s) describing the supermembrane contribution(s) to the supergravity (super)field equations, one needs to vary the Wess--Zumino term of the supermembrane action with respect to the supergravity (super)fields. Thus one arrives at a separate problem of either constructing the 3-form potential from the supergravity pre-potential or to find the variation
\begin{eqnarray} \label{vC3:=}
 \delta C_3= {1\over 3!} E^C\wedge E^B\wedge E^A \beta_{ABC}(\delta)\;    \qquad
\end{eqnarray}
such that $d\delta C_3=\delta H_4$ reproduces the variation of $H_4$ written in terms of
the basic supergravity variations (we describe these in the next section).

Studying such a technical problem we found that it imposes a restriction on the independent variations of the supergravity prepotentials, or equivalently, on the independent parameters of the admissible supervielbein variations,  thus transforming the generic minimal supergravity into a {\it special minimal supergravity}.
This off--shell supergravity formulation had been described for the first time in \cite{Ovrut:1997ur} using the elegant combination of superfield results and the component 'tensor calculus' approach on the line of \cite{BW}.
We describe this {\it special minimal supergravity}  below, in the complete Wess--Zumino superfield formalism, after presenting briefly some  necessary technical details on generic minimal supergravity.

\subsection{Superfield supergravity action and superfield supergravity equations. }
\label{SGacSGeq}

\medskip 

It is well known that superfield supergravity action is given by the volume of the superspace \cite{WZ78},  this is to say by integral over the curved supergravity superspace
$\Sigma^{(4|4)}$ of the Berezenian (superdeterminant) $E:= sdet(E_M^A)$ of the supervielbein $E_M^A(Z)$,
\begin{eqnarray}\label{SGact} S_{SG} = \int d^4 x \tilde{d}^4\theta \; sdet(E_M^A) \; \equiv \int d^8Z \;
E \; .
\end{eqnarray}
Indeed, assuming that the supervielbein superfields $E_M^A(Z)$ are subject to the constraints (\ref{4WTa=}), (\ref{4WTal=}) and performing the Berezin integration over the Grassmann coordinate of  $\Sigma^{(4|4)}$, Wess and Zumino arrived \cite{WZ79} at the spacetime component supergravity action with minimal set of auxiliary fields \cite{TenCal}.

The equations of motion of supergravity can be obtained from such a component action\footnote{Such a  component action for the 'special minimal supergravity' was presented in \cite{Ovrut:1997ur}. Below we will be following another way oriented on derivation of superfield equations by varying the superfield action (\ref{SGact}. }. The superfield supergravity equations can be found knowing these spacetime component equations and the consequences of the off--shell supergravity constraints in Eqs. (\ref{4WTa=}), (\ref{4WTal=}) which express the superfield generalization of the {\it l.h.s.}'s of these equations of motion in terms of the off--shell supergravity superfields. Namely (see \cite{BW} and \cite{BdAIL03}), as a consequence of the constraints   (\ref{4WTa=}), (\ref{4WTal=}) the superfield generalization of the {\it l.h.s.} of the supergravity Rarita--Schwinger equation reads
\begin{eqnarray}\label{SGRS=off}
\epsilon^{abcd}T_{bc}{}^{\alpha}\sigma_{d\alpha\dot{\alpha}} ={i\over 8}
\tilde{\sigma}^{a\dot{\beta}\beta} \bar{{\cal D}}_{(\dot{\beta}|} G_{\beta|\dot{\alpha})} + {3i\over 8}
{\sigma}^a_{\beta \dot{\alpha}} {\cal D}^{\beta}R \; ,
\end{eqnarray}
and the superfield generalization of the Ricci tensor is
\begin{eqnarray} \label{RRicci}
R_{bc}{}^{ac}& = {1\over 32} ({{\cal D}}^{{\beta}} \bar{{\cal D}}^{(\dot{\alpha}|}
G^{{\alpha}|\dot{\beta})} - \bar{{\cal D}}^{\dot{\beta}} {{\cal D}}^{({\beta}}G^{{\alpha})\dot{\alpha}})
\sigma^a_{\alpha\dot{\alpha}}\sigma_{b\beta\dot{\beta}} - {3\over 64}
(\bar{{\cal D}}\bar{{\cal D}}\bar{R} + {{\cal D}}{{\cal D}}{R}- 4 R\bar{R})\delta_b^a\; .
\end{eqnarray}
This suggests that superfield supergravity equation should have the form
 \begin{eqnarray}\label{SGeqmG}
 G_a =0 \; ,
\\ \label{SGeqmR}
 R=0 \; ,
\end{eqnarray}
and c.c. to Eq. (\ref{SGeqmR}). Indeed, substituting these in  Eqs. (\ref{SGRS=off}) and (\ref{RRicci}) we find the superfield generalization of the Rarita--Schwinger and Einstein equations of supergravity,
\begin{eqnarray}
\label{SGRS=0}
&& \epsilon^{abcd}T_{bc}{}^{\alpha}\sigma_{d\alpha\dot{\alpha}} = 0 \; , \qquad
\\ \label{RRicci=0}
&& R_{bc}{}^{ac} = 0 \; .
\end{eqnarray}

\subsection{Admissible variation of constrained supervielbein and prepotential structure of generic minimal supergravity.}

One can also obtain the supergravity superfield  equations (\ref{SGeqmG}), (\ref{SGeqmR}) by varying the superspace action (\ref{SGact}).  This is not apparent because the supervielbein superfields are restricted by the constraints (\ref{4WTa=}), (\ref{4WTal=}). There are two basic ways to solve this problem. The first consists in solving the  superspace supergravity constraints in terms of unconstrained superfields-- {\it pre-potentials} \cite{OS78,Siegel:1978nn,Siegel79,OS80}  -- the set of which in the case of minimal supergravity can be  restricted to the axial vector superfield ${\cal H}^\mu (Z)= {\cal H}^\mu (x,\theta,\bar{\theta})$ \cite{OS78}  and the chiral compensator superfield $\Phi(x+i{\cal H}(Z), \theta)$ \cite{Siegel:1978nn}. Using this in our present study  would imply finding the expression for the 3--form $C_3$ in terms of ${\cal H}^\mu (Z)$, $\Phi(x+i{\cal H}(Z), \theta)$ and $\bar{\Phi}(x-i{\cal H}(Z), \theta)=(\Phi(x+i{\cal H}(Z), \theta))^*$.

Another way, called the Wess--Zumino approach to superfield supergravity \cite{WZ77,GWZ78,WZ78,WZ79,BW}, does not imply to solve the constraints but rather to solve the equations which appear as a result of requiring that the variations of the supervielbein and spin connection
\begin{eqnarray}\label{varEMA} \delta
E_M^{\, A}(Z) = E_M^{\, B} {\cal K}_{B}^{\, A} (\delta )\; , \quad \delta w_M^{ab}(Z) = E_M^{\, C}
u_{{C}}^{ab} (\delta )\; , \end{eqnarray}
preserve the superspace supergravity constraints  \cite{WZ78}, Eqs. (\ref{4WTa=}), (\ref{4WTal=}).

Actually this procedure of finding {\it admissible} variations of the supervielbein and superspace spin connection is a linearized but covariantized version of the constraint solution used in pre-potential approach. The independent parameters of admissible variations clearly reflect the pre-potential structure of the off-shell supergravity. In the case of minimal supergravity their set contain \cite{WZ78} (see also \cite{BdAIL03}) $\delta H^a$, corresponding to the variation of the axial vector pre-potential of \cite{OS78}, and complex scalar variations $\delta {\cal U}$  entering  (\ref{varEMA}) only under the action of chiral projectors (as $({\cal D}{\cal D}-\bar{R})\delta {\cal U}$) and thus corresponding to the variations of complex pre-potential of the chiral compensator of the minimal supergravity (as well as   $\delta \bar{{\cal U}}=(\delta {\cal U})^*$ entering through $(\bar{\cal D}\bar{\cal D}-{R})\delta \bar{{\cal U}}$).
The admissible variations of supervielbein read \cite{WZ78,BdAIL03}
\begin{eqnarray}\label{varEa} \delta E^{a} & =& E^a (\Lambda (\delta ) + \bar{\Lambda} (\delta ))
 - {1\over 4} E^b \tilde{\sigma}_b^{ \dot{\alpha} {\alpha} }
[{\cal D}_{{\alpha}}, \bar{{\cal D}}_{\dot{\alpha}}] \delta H^a +
 i E^{\alpha} {\cal D}_{{\alpha}}\delta H^a   - i \bar{E}^{\dot{\alpha}}\bar{{\cal
D}}_{\dot{\alpha}} \delta H^a \; , \\ \label{varEal}
 \delta E^{\alpha} & = & E^a \Xi_a^{\alpha}(\delta ) +
E^{\alpha} \Lambda (\delta ) + {1\over 8} \bar{E}^{\dot{\alpha}} R \sigma_a{}_{\dot{\alpha}}{}^{\alpha}
\delta H^a \; ,  \end{eqnarray}
where
\begin{eqnarray}
\label{2Lb+*Lb} & 2\Lambda (\delta ) + \bar{\Lambda} (\delta )  = {1\over 4} \tilde{\sigma}_a^{
\dot{\alpha} {\alpha} } {\cal D}_{{\alpha}} \bar{{\cal D}}_{\dot{\alpha}}\delta H^a + {1\over 8} G_a
\delta H^a +   3 ( {\cal D}{\cal D}- \bar{R})\delta {\cal U} \;
 \end{eqnarray}
and the explicit expression for $\Xi_a^{\alpha}(\delta )$ in (\ref{varEal}) can be found in \cite{BdAIL03}. Neither that  nor the explicit expressions for the admissible variations of spin connection in (\ref{varEMA}) will be needed for our discussion below.

Indeed, the variation of superdeterminant of the supervielbein of the minimal supergravity superspace can be calculated using (\ref{varEa}), (\ref{varEal}) only and reads
 (see \cite{WZ78})
\begin{eqnarray}\label{varsdE}
\delta E = & E [ - {1\over 12} \tilde{\sigma}_a^{\dot{\alpha}\alpha} [ {\cal D}_{\alpha}, \bar{\cal
D}_{\dot{\alpha}}] \delta H^a + {1\over 6} G_a \; \delta H^a +
  2(\bar{\cal D} \bar{\cal D} - R) \delta \bar{{\cal U}} +  2({\cal D} {\cal D} - \bar{R}) \delta {\cal
U}] \; . \end{eqnarray}
Taking into account the identity \cite{WZ78}
\begin{eqnarray} \label{idd80} \int d^8Z\,   E \; ({\cal D}_{A}\xi^{A} + \xi^B T_{BA}{}^A) (-1)^{A}\; \equiv 0 \; ,
\end{eqnarray}  one  finds the variation  of the minimal supergravity action  (\ref{SGact})
 \begin{eqnarray}\label{vSGsf} & \delta S_{SG} =  \int
d^8Z E\;  [{1\over 6} G_a \; \delta H^a -  2 R\; \delta \bar{{\cal U}} -2 \bar{R}\; \delta {\cal U}] \; . \quad
\end{eqnarray}
This clearly produces the superfield supergravity equations of the form (\ref{SGeqmG}), (\ref{SGeqmR}) which result in the Rarita--Schwinger equation (see (\ref{SGRS=0})) and
Einstein equation without cosmological constant (see (\ref{RRicci=0})).

To obtain Einstein equation with cosmological constant one can add a specific term (homogeneous of order three in chiral compensator, see p. 336 of \cite{1001} and p. 682 of \cite{SiBook}). But here we will be rather interested in  a dynamical generation of the cosmological constant, as this is characteristic for the   {\it special minimal supergravity} and for the Ogievetski--Sokatchev theory of axial vector superfield, which are based on the standard supergravity action (\ref{SGact}), and allow for a dynamical coupling to supermembrane.

\section{Variant superfield equations and Ogievetsky--Sokatchev mechanism of the dynamical generation of cosmological constant.}
\label{OSdg}

\medskip 

Coming back to the discussion in Sec.
\ref{SGacSGeq}, let us observe, following  \cite{OS80}, that, if we impose only vector superfield equation, Eq. (\ref{SGeqmG}), then, as a result of (\ref{DG=DR}), the complex scalar main superfield $R$, which is chiral due to (\ref{chR}), also obeys ${\cal D}_\alpha R=0$; hence it becomes equal to a (complex) constant,
\begin{eqnarray} \label{SGeqmR=OS}
 R=c_1+ic_2 \; , \qquad  \bar{R}=c_1-ic_2 \; , \qquad c_{1,2}= const\; . \;
\end{eqnarray}
This does not change the fermionic equation (\ref{SGRS=0}), but modifies the Einstein equation by a cosmological constant contribution,
\begin{eqnarray}
 \label{RRicci=c12+c22}
& R_{bc}{}^{ac}  = {3\over 16} (c_1^2+c_2^2) \, \delta_b{}^a\; .
\end{eqnarray}
The cosmological constant  $-\Lambda= {3\over 8} (c_1^2+c_2^2) $ is expressed through the arbitrary constants $c_1$ and $c_2$ (integration constants) so that one can state that it is {\it generated dynamically}.

This mechanism of cosmological constant generation was first observed by Ogievetski and Sokatchev \cite{OS80} in their formulation of supergravity as a theory of axial gravitational superfield \cite{OS78}.
Instead of the introduction of the chiral compensator, which appears when solving supergravity constraints  \cite{Siegel:1978nn}, Ogievetski and Sokatchev restricted the relevant supergroup of conformal supergravity, as described by axial vector superfield.
The result is that, instead of two scalar auxiliary fields of minimal supergravity, their theory of  axial vector superfield contains divergences of two auxiliary vector fields (which enter the theory under divergences only). This gives rise to the mechanism of cosmological constant generation which is similar (although not identical) to the one  discussed in \cite{CCgen}  in terms of componet fields (see also \cite{CCgenN8}) and also to the  mechanism of dynamical generation of tension of $p$--branes studied in \cite{TenGen=PKT92,TenGen=BLT92}.

In the Wess--Zumino approach the Ogievetski--Sokatchev formulation of supergravity can be formally described by suppressing (by a brut force) the variations $\delta {\cal U}$, $\delta \bar{{\cal U}}$ corresponding to the chiral compensator. Then the supergrvaity action variation (\ref{vSGsf}) reduces to  $\delta S_{SG} = {1\over 6} \int
d^8Z E\;   G_a \; \delta H^a $ and the only superfield equation is $G_a=0$, as above. Presently it is unclear how to obtains such restriction of the basic supergravity variations starting from the solution of the Wess--Zumino torsion constraints
\footnote{The supergroup of conformal supergravity can be used to fix the gauge where the chiral compensator is equal to, say, unity. In such a way, starting from the solution of the constraints of minimal supergravity \cite{Siegel:1978nn}, one arrives at the Ogievetsy--Sokatchev formalism. One can notice a similarity of the effect of missing, after such a gauge fixing, of  one superfield equation $R=0$, which we have reviewed above, and of the missing Virasoro constraints in the bosonic string theory, when fixing the conformal gauge. Notice that in both cases, the missing of equations occurs when the gauge fixed symmetry is conformal symmetry. }.

However, if we restrict the set of minimal supergravity variation by expressing $\delta {\cal U}$ in terms of {\it real variation} $\delta V= (\delta V)^*$, namely (and schematically) put  $\delta {\cal U}=i \delta V$ instead of setting it to zero, then, on one hand, the effect of dynamical generation of cosmological constant still occurs and, on the other hand, such a restriction appears naturally in the Wess-Zumino formalism. Namely, the requirement of the existence of the superspace three--form potential $C_3$ constructed from the supergravity superfields does restrict the set of basic variations of minimal supergravity by the condition described above. In other words, this requirement modifies the auxiliary field content of the supergravity: in  {\it special minimal supergravity} allowing for the existence of $C_3$ constructed from the supergravity superfields the chiral compensator superfield does exist, but is expressed in terms of {\it real} superfield pre-potential, rather than in terms of complex scalar prepotential as it is the case for generic chiral superfield. 

Let us repeat that such a 'special minimal supergravity' was described for the first time by Ovrut and Waldram \cite{Ovrut:1997ur} under the name of 'three form supergravity' using component approach  together with elements of superfield approach to three--form matter interacting with supergravity from \cite{Binetruy:1996xw}. The dynamical generation of cosmological constant in special minimal supergravity was described in \cite{Ovrut:1997ur} in the language of  component approach.
The prepotential approach to special minimal supergravity was discussed in \cite{Kuzenko+05} starting just form imposing on the chiral compensator of minimal supergravity the condition to be the special minimal superfield constructed from real prepotential.
\footnote{The study of supermembrane interaction with dynamical scalar multiplet also resulted in the conclusion that, the supermembrane allows the coupling to dynamical scalar multiplet iff this is described by special scalar superfield, expressed through real prepotential \cite{IB+CM=2010}. This is in agreement with the results of early study of variant superfield representation in \cite{Gates:1980ay,Gates:1980az} where the expression for 4-form field strength in flat superspace was found in terms of arbitrary scalar superfield, but the expression for the 3-form potential was given only for the special scalar superfield constructed from the real prepotential.}.

\section{Special minimal supergravity.}

\medskip 

\subsection{Three form gauge potential in the superspace of minimal supergravity requires modifying its auxiliary field sector.}

\medskip 

Using the supergravity variations (\ref{varEa}), (\ref{varEal}) one can calculate the variation of the closed 4--form (\ref{H4=HL+HR}),
\begin{eqnarray} \label{vH4=}  \delta H_4 &=& {1\over 2} E^b\wedge  E^\alpha \wedge  E^\beta\wedge  E^\gamma
\sigma_{ab\; (\alpha\beta} D_{\gamma )}\delta H^a  - {1\over 2} E^b\wedge  E^\alpha \wedge  E^\beta\wedge   \bar{E}{}^{\dot{\gamma}}
\sigma_{ab\; \alpha\beta} \bar{D}_{\dot\gamma}\delta H^a + c.c. - \qquad
 \nonumber \\ && - {i\over 2} E^b\wedge E^a \wedge E^\alpha \wedge E^\beta
\left(\sigma_{ab\; \alpha\beta}\left(2\Lambda(\delta) + \bar{\Lambda}(\delta)\right)+ {1\over 4}\sigma_{c[a|\; \alpha\beta}\tilde{\sigma}_{|b]}{}^{\dot{\gamma}\gamma}[D_\gamma, \bar{D}_{\dot{\gamma}}]\delta H^c
\right) + c.c. + \nonumber \\ && + {i\over 16} E^b\wedge  E^a \wedge  {E}{}^{\alpha} \wedge \bar{E}{}^{\dot\beta} (R\sigma_{ab}\tilde{\sigma}_{c} -\bar{R}\sigma_c\tilde{\sigma}_{ab})_{\alpha\dot\beta}\delta H^c + \propto  \; E^c \wedge E^b\wedge E^a  \; .
\end{eqnarray}
On the other side, if we assume that $H_4=dC_3$, we should be able to express
$\delta H_4$ in terms of the variation of the 3--form potential $\delta C_3$,
 \begin{eqnarray} \label{vH4=dvC3}
\delta H_4 = d(\delta C_3)
 \; , \quad
\end{eqnarray}
and may assume that $\delta C_3$ is decomposed on the basic covariant 3--forms, as in Eq. (\ref{vC3:=}).

Clearly, the gauge transformations $ \delta C_3=d \alpha_2$ with $\alpha_2= {1\over 2}  E^B\wedge E^A \alpha_{AB}$ do not change the field strength and thus are not of any interest at the present stage. If interested in $\delta C_3$ variations which produce $\delta H_4$ in (\ref{vH4=}), we can  simplify the general expression for $\delta C_3$ variation in (\ref{vH4=dvC3}) by factoring out the gauge transformations. Particularly, we can start with $\delta C_3$  of the form (\ref{vC3:=}) with
\begin{eqnarray} \label{bff*f=}
\beta_{\alpha\beta\gamma}(\delta)=0= \beta_{\alpha\beta\dot{\gamma}}(\delta)\; , \quad \beta_{{\alpha}\dot{\beta} a}(\delta)  = i\sigma_{a\alpha\dot{\beta}}\delta V \; .
\end{eqnarray}
Then, after straightforward but a bit involving  calculations, one finds
\begin{eqnarray} \label{bffb=}
\beta_{\alpha\beta a}(\delta) &=&  - \sigma_{ab\; \alpha\beta} (\delta H^b + \tilde{\sigma}^{b\gamma \dot{\gamma}}D_{\gamma }\delta\bar{\kappa}_{\dot{\gamma}})\; ,    \qquad
 \\ \label{bfbb=}
\beta_{\alpha ab}(\delta) &=& {1\over 2}\epsilon_{abcd}\sigma^c_{\alpha\dot\alpha}\bar{D}{}^{\dot\alpha }\delta H^d +{1\over 2} \sigma_{ab\; \alpha}{}^{\beta} D_\beta \delta V  -
{i\over 4} \tilde{\sigma}_{ab}{}^{\dot{\beta}}{}_{\dot{\gamma}} \bar{D}_{\dot\beta}D_\alpha \bar{\kappa}{}^{\dot{\gamma}}  + {i\over 4} \sigma_{ab\; \alpha}{}^{\beta} \bar{D}_{\dot\beta}D_\beta \bar{\kappa}{}^{\dot{\beta}}   \; ,    \qquad \\
\label{bbbb=}
\beta_{abc}(\delta)  &=& {i\over 8} \epsilon_{abcd}\left( \left({\bar{\cal D}}{\bar{\cal D}}-
{1\over 2}R \right)\delta H^d - c.c. \right) + \nonumber \quad \\   &+& {1\over 4} \epsilon_{abcd}G^d\delta V + {1\over 8} \epsilon_{abcd}\tilde{\sigma}{}^{d\dot{\gamma}\gamma}
[{\cal D}_{\gamma}, {\bar{\cal D}}_{\dot{\gamma}}]\delta V
 - {i\over 16}\epsilon_{abcd}\tilde{\sigma}{}^{d\dot{\gamma}\gamma}\left(\left({\cal D}{\cal D}+{5\over 2}\bar{R}\right) {\bar{\cal D}}_{\dot{\gamma}} {\kappa}_{{\gamma}}- c.c.
  \right)\, .  \qquad
\end{eqnarray}

Furthermore, the consistency between Eq. (\ref{vH4=}) and Eqs. (\ref{vC3:=}), (\ref{bff*f=}), (\ref{bffb=}), (\ref{bfbb=}) requires that the complex scalar variation $\delta {\cal U}$ (or, better to say, $({\cal D}{\cal D}-\bar{R})\delta {\cal U}$) is expressed in terms of real $\delta V$ and $\bar{D}_{\dot{\alpha}} \delta\bar{\kappa}^{\dot{\alpha}}$  by  (the solution of)
\begin{eqnarray}
 \label{DDcU=vV+}
({\cal D}{\cal D}- \bar{R})\delta {\cal U} = {1\over 12}( {\cal D}{\cal D}- \bar{R})\left(i\delta {V} + {1\over 2}  \bar{{\cal D}}_{\dot{\alpha}}  \delta \bar{{\kappa}}{}^{\dot{\alpha}} \right)
\; .    \;
\end{eqnarray}
Eq. (\ref{DDcU=vV+}) is equivalent to
\begin{eqnarray}
 \label{vcU=vV+}
\delta {\cal U} = {i\over 12}\delta {V} + {1\over 24}  \bar{{\cal D}}_{\dot{\alpha}}  \delta \bar{{\kappa}}{}^{\dot{\alpha}}+ {i\over 24} {\cal D}_{{\alpha}} \delta {{\nu}}{}^{{\alpha}}\; ,   \qquad
\end{eqnarray}
where  $\delta {{\nu}}{}^{{\alpha}}$ is an additional independent  variation; this however does not contribute to $({\cal D}{\cal D}-\bar{R})\delta {\cal U}$ and, hence, to the variations of supergravity potentials.

Thus, although the off--shell superspace of generic minimal supergravity allows for the existence of closed 4-form (\ref{H4=HL+HR}), (\ref{H4L}), the requirement of the existence of the corresponding 3-form potential $C_3$ with the variation (\ref{vC3:=}), such that $H_4=dC_3$, modifies the set of basic variations of supergravity. That still contains unrestricted $\delta H^a$, but as far as the complex variation $\delta {\cal U}$ and $\delta \bar{{\cal U}}=(\delta {\cal U})^*$ are concerned, only their imaginary part $i\delta V$ remains independent. This is the reformulation of the statements  of \cite{Ovrut:1997ur} and \cite{Kuzenko+05} on the field content of 'special minimal supergravity' (or three form supergravity) in the language of the basic variations of superfield supergravity.

To be precise, above we should also mention the variations $\delta\bar{\kappa}^{\dot{\alpha}}$ entering (\ref{DDcU=vV+}) in the form of $\bar{{\cal D}}_{\dot{\alpha}}  \delta \bar{{\kappa}}{}^{\dot{\alpha}}$. As we will see in a moment, $\delta\bar{\kappa}^{\dot{\alpha}}$ and its {\it c.c.} $\delta{\kappa}^{{\alpha}}$ describe symmetries of the superfield supergravity action (\ref{SGact}) and in this sense are inessential when pure supergravity theory is considered. The question whether it is a symmetry of the supergravity---supermembrane interacting system remains open\footnote{The component results in \cite{Ovrut:1997ur} suggest that this is the case.}; we leave it for the forthcoming paper devoted to the detailed study of this dynamical system \cite{IB+CM=2011}.

The above modification of the set of basic supergravity variations ($(\delta {\cal U}, \delta \bar{{\cal U}})  \mapsto \delta V=(\delta V)^*$) corresponds to that the auxiliary fields come now from a {\it special chiral superfield}, expressed through the real prepotential, rather than from the usual chiral superfield expressed through the complex prepotential (see \cite{IB+CM=2010} for the description of a special chiral multiplet described by such a type of chiral matter superfield, with $D=4$ ${\cal N}=1$ supermembrane). This also imply that, effectively, one of the scalar auxiliary fields of minimal supergravity is substituted by a divergence of an auxiliary vector field $\partial_\mu k^\mu$, or,  equivalently by $\epsilon^{\mu\nu\rho\sigma}\partial_\mu B_{\mu\rho\sigma}$, as in the original description of this {\it special minimal supergravity} by Ovrut and Waldram \cite{Ovrut:1997ur}. 

\subsection{Superfield equations  and  dynamical generation of cosmological constant in special minimal supergravity.}

\medskip 

In the Wess--Zumino approach, the action of special minimal supergravity is given by the same supervolume formula, Eq. (\ref{SGact}), but the variations of supervielbein entering this action are now restricted by the conditions of not only preserving the supergravity constraints, (\ref{4WTa=}), (\ref{4WTal=}), but also reproducing the closed four form variation (\ref{vH4=}) from the variation of the three form potential  (\ref{vH4=dvC3}) of the form (\ref{vC3:=}). This is to say the variations of the supervielbein are given by Eqs. (\ref{varEa}), (\ref{varEal}) but with $\delta {\cal U}$ given by
Eqs. (\ref{vcU=vV+}) or, equivalently,   by Eq. (\ref{DDcU=vV+}).

The variation of the superdeterminant is then given by ({\it cf.} Eq. (\ref{varsdE})) \begin{eqnarray}\label{varsdE=s}
\delta E &= E \left[-{1\over 12}  \tilde{\sigma}_a^{\dot{\alpha}\alpha} [ {\cal D}_{\alpha}, \bar{\cal
D}_{\dot{\alpha}}] \delta H^a + {1\over 6} G_a  \delta H^a
  - \left({1\over 6} ( {\cal D}{\cal D}- \bar{R})\left(\delta {V}
- {1\over 2}  \bar{{\cal D}}_{\dot{\alpha}}  \delta \bar{{\kappa}}{}^{\dot{\alpha}} \right) + c.c. \right)\right]
\; 
\end{eqnarray}
and the variation of the special minimal supergravity action reads 
\begin{eqnarray}\label{vSGsf=s}  \delta S_{SG} &=&  {1\over 6} \int
d^8Z E\;  \left[ G_a \; \delta H^a + (R-\bar{R}) i\delta {V}  \right] - \nonumber \\ &&
 - {1\over 12} \int
d^8Z E\; \left(R {\cal D}_{{\alpha}} \delta {{\kappa}}{}^{{\alpha}}  +
\bar{R} \bar{{\cal D}}_{\dot{\alpha}}  \delta \bar{{\kappa}}{}^{\dot{\alpha}}\right)
 \; . \qquad
\end{eqnarray}

Eq. (\ref{vSGsf=s}) still produces the vector superfield equation (\ref{SGeqmG}),
\begin{eqnarray}\label{SGeqmG=0}
G_a=0\; , \qquad
\end{eqnarray}
but instead of the complex scalar superfield equations (\ref{SGeqmR}) valid in the case of generic minimal supergravity, in the case of special minimal supergravity we have only the real scalar equation
\begin{eqnarray}\label{SGeqmR+*=0}
R- \bar{R}=0\; . \qquad
\end{eqnarray}
Clearly, due to chirality of $R$, $\;\bar{{\cal D}}_{\dot\alpha} {R}=0$,  and anti-chirality of $\bar{R}$, $\; {\cal D}_\alpha \bar{R}=0$ (see Eqs. (\ref{chR})), the above Eq. (\ref{SGeqmR+*=0}) also implies that $\bar{{\cal D}}_{\dot\alpha}\bar{R}=0$ and ${\cal D}_\alpha {R}=0$.  Using this and the algebra of covariant derivatives,
$\{ {\cal D}_\alpha ,{\cal D}_{\dot\alpha}\} = 2i \sigma^a_{\alpha {\dot\alpha}}{\cal D}_a $, we find  that, as a result of (\ref{SGeqmR+*=0})
the complex superfield $R$ is actually equal to a {\it real}  constant,
\begin{eqnarray}\label{bR=4ic}
R=4c \; , \qquad \bar{R}=4c\; , \qquad c=const=c^*\; .
\end{eqnarray}
(the coefficient $4$ is introduced into (\ref{bR=4ic}) for future convenience).

Now, using (\ref{RRicci}), we observe that the superfield equation (\ref{SGeqmR+*=0}) results in Einstein equation with cosmological constant
\begin{eqnarray}\label{RRici=c}
R_{bc}{}^{ac}= 3c^2 \delta_b{}^a\; .
\end{eqnarray}
The value of the cosmological constant is proportional to the square of the above arbitrary constant $c$, which has appeared as an integration constant, so that the special minimal supergravity is characterized by a {\it cosmological constant generated dynamically}.

The above mechanism of the dynamical generation of cosmological constant in special minimal supergravity is the same as was observed by  Ogievetski--Sokatchev \cite{OS80} in their theory of axial vector superfield. Despite we have  already  described this mechanism  in Sec. \ref{OSdg}, we have found useful to discuss in detail how it works in special minimal supergravity, the off--shell supergravity formulation which can be used to calculate the supermembrane current.
In the language of component spacetime approach to supergravity the dynamical generation of cosmological constant in special minimal supergravity was described already in \cite{Ovrut:1997ur}.

This is the place to notice that the variations $\delta {\kappa}^{\alpha}$ produce the equation ${\cal D}_\alpha R=0$, which is not independent: as we have discussed above, they follow form Eq. (\ref{bR=4ic}) and (\ref{chR}). Hence these variations describe the gauge symmetry of the special minimal supergravity action. As we have already said above, we postpone the discussion on the $\delta {\kappa}^{\alpha}$ transformations  in the supergravity---supermembrane interacting system till the forthcoming paper \cite{IB+CM=2011}.

\section{Supermembrane supercurrent  and its contribution to the supergravity superfield equations.}

\medskip 
Now we see that the variation of the supermembrane action (\ref{Sp=2:=}) with respect to the vector prepotential of supergravity gives us the vector supercurrent of the form
\begin{eqnarray}\label{Ja=NG+WZ}
J_a&=&
\int\limits_{W^3} {1\over 2\hat{E}} \hat{E}{}^b \wedge \hat{E}{}^\alpha \wedge \hat{E}{}^\beta\; {\sigma}_{ab \alpha\beta}  \delta^8 (Z-\hat{Z}) - \qquad \nonumber
\\ && -
\int\limits_{W^3} {i\over 2\hat{E}} \left( *\hat{E}_a \wedge \hat{E}{}^\alpha - {i\over 2}   \hat{E}{}^b \wedge \hat{E}{}^c \wedge \hat{\bar{E}}_{\dot{\beta}}\epsilon_{abcd} \tilde{\sigma}^{d\dot{\beta}\alpha}\right) {\cal D}_\alpha \delta^8 (Z-\hat{Z})
+c.c +
\nonumber
\\
&& +\int\limits_{W^3}  {1\over 2\cdot 4!\hat{E}}   \, \hat{E}{}^b \wedge \hat{E}{}^c \wedge \hat{E}{}^d \, \epsilon_{abcd} \left( {\cal D}{\cal D}- {1\over 2}\bar{R} \right) \delta^8 (Z-\hat{Z}) + c.c.  + \qquad \nonumber
\\ && + \int\limits_{W^3}  {1\over 4!\hat{E}} *\hat{E}_b \wedge \hat{E}{}^b  \, G_a\,  \delta^8 (Z-\hat{Z})  -  \nonumber  \qquad
\\
&& -\int\limits_{W^3}  {1\over 4!\hat{E}}\; *\hat{E}_c \wedge \hat{E}{}^b \tilde{\sigma}^{d\dot{\alpha}\alpha} \left( 3\delta_a^c \delta_b^d-  \delta_a^d \delta_b^c \right)[{\cal D}_\alpha , \bar{\cal D}_{\dot{\alpha}}] \delta^8 (Z-\hat{Z})  \; , \quad
\end{eqnarray}
where ${\hat{E}}= {sdet (E_M{}^A(\hat{Z}))}$ and
\begin{eqnarray}\label{d8:=}
 \delta^8(Z):={1\over 16}\,\delta^4(x) \, \theta\theta \, \bar{\theta}\bar{\theta}\; , \qquad \;
\end{eqnarray}
is the superspace delta function which obeys $\int d^8Z\, \delta^8(Z-Z') f(Z) = f(Z')$ for any superfield $f(Z)$.

The supercurrent (\ref{Ja=NG+WZ}) enters the {\it r.h.s.} of the vector superfield equation
\begin{eqnarray}\label{Ga=Ja}
 G_a=J_a\;
\end{eqnarray}
which follows from the action of the supergravity---supermembrane interacting system
\begin{eqnarray}\label{Sint=SG+Sp2}
 S&=& S_{SG}+ {1\over 6}S_{p=2}= \int d^8 Z E(Z) +
 {1\over 12}\int d^3 \xi \sqrt{g} - {1\over 6} \int\limits_{W^3} \hat{C}_3\; ,
\end{eqnarray}
(as $\delta S/\delta H^a=0$; the coefficient ${1\over 6}$ introduced in (\ref{Sint=SG+Sp2}) for convenience).
The scalar superfield equation of the interacting system, which is obtained by varying the interacting action (\ref{Sint=SG+Sp2}) with respect to  the real scalar prepotential of special minimal supergravity, $\delta S/\delta V=0$,  reads
\begin{eqnarray}\label{R-bR=cX}
R-\bar{R}= -i {\cal X}
\end{eqnarray}
where
\begin{eqnarray}\label{cX=}
{\cal X} =  {i\over {E}}  \int\limits_{W^3} \hat{E}^a \wedge \hat{E}{}^\alpha  \wedge \hat{\bar{E}}^{\dot{\alpha}}\, {\sigma}^a_{\alpha\dot{\alpha}}\; \delta^8 (Z-\hat{Z}) -
\nonumber  \qquad
\\  - \int\limits_{W^3} { \hat{E}{}^b \wedge \hat{E}{}^a \wedge \hat{E}{}^\alpha \over 4\hat{E}}   \; {\sigma}_{ab \alpha}{}^{{\beta}} {\cal D}_{\beta} \delta^8 (Z-\hat{Z})
+c.c + \nonumber  \;
\\ +\int\limits_{W^3}  {   \, \hat{E}{}^b \wedge \hat{E}{}^c \wedge \hat{E}{}^d \over 2\cdot 4!\hat{E}}\, \epsilon_{abcd} \tilde{\sigma}^{a\dot{\alpha}\alpha} [{\cal D}_\alpha , \bar{\cal D}_{\dot{\alpha}}] \delta^8 (Z-\hat{Z})+\;
\nonumber  \;
\\ + i\int\limits_{W^3}  {*\hat{E}_a \wedge \hat{E}{}^a \over 4!\hat{E}}
\left( {\cal D}{\cal D}- \bar{R} \right) \delta^8 (Z-\hat{Z}) + c.c. \;  \qquad \nonumber \;
\\+\int\limits_{W^3}  {1 \over 4!\hat{E}} \hat{E}{}^b \wedge \hat{E}{}^c \wedge \hat{E}{}^d \epsilon_{abcd} G{}^{a}\; \delta^8 (Z-\hat{Z})\; . \qquad
\end{eqnarray}

Notice that, as a consequence of (\ref{DG=DR}), the supermembrane current superfields obey
\begin{eqnarray}\label{DJ=-iDX}
\bar{{\cal D}}{}^{\dot{\alpha}} J_{\alpha\dot{\alpha}} = i {{\cal D}}_{{\alpha}} {\cal X}
\; , \qquad
{{\cal D}}{}^{{\alpha}} J_{\alpha\dot{\alpha}} = -i \bar{{\cal D}}_{\dot{\alpha}}{\cal X}\; .
\end{eqnarray}
Although at first glance these relations look different from any of listed in \cite{Seiberg+2009,Kuzenko:2010}, they can be reduced to the Ferrara--Zumino multiplet if one takes into account Eq. (\ref{R-bR=cX}). Indeed, this states that the real superfield ${\cal X}$ in the {\it r.h.s.} of Eq. (\ref{DJ=-iDX}) is the sum of chiral superfield (equal to $iR$) and its complex conjugate, so that only the first (second) one contributes to the {\it r.h.s.} of the first (second) equation in (\ref{DJ=-iDX}). Certainly, the statement of being the sum of chiral superfield and its complex conjugate is a selfconsistency condition imposed on the real scalar superfield (\ref{cX=}); the other consistency conditions are
provided by Eq. (\ref{DJ=-iDX}) and also by vanishing of the fermionic currents $\delta S_{p=2}/\delta \kappa^\alpha$ on the mass shell of the interacting system. All these will be studied in the forthcoming paper \cite{IB+CM=2011} devoted to the detailed study of the properties of the supergravity--supermembrane interacting system described by the superfield action (\ref{Sint=SG+Sp2}).

The explicit form of supercurrent superfields (\ref{Ja=NG+WZ}), (\ref{cX=}) and of the supergravity superfield equations with supermembrane contributions are  the main results of the present paper.

\section{Conclusions and discussion.}

In this contribution we have derived the explicit form of supercurrent superfields of the $D=4$, ${\cal N}=1$ supermembrane and show how these enter the superfield supergravity equations.

To this end we had to find an appropriate off--shell supergravity formulation which allowed  the coupling to supermembrane, so that a significant part of the paper is devoted to the issues of $D=4$, ${\cal N}=1$  supergravity.
Such an off--shell formulation of supergravity, which we prefer to call 'special minimal supergravity', had been found by Ovrut and Waldram \cite{Ovrut:1997ur} (who called it 'three form supergravity'). One can arrive at this formulation
studying the consequences of the existence of a three form gauge potential in the superspace subject to minimal supergravity constraints. Although the generic (off-shell) minimal supergravity superspace allows for the existence of closed 4-form $H_4$ (see Eqs. (\ref{H4=HL+HR}), (\ref{H4L})), the requirement of the existence of the corresponding 3-form potential $C_3$ with the variation (\ref{vC3:=}), such that $H_4=dC_3$, modifies the set of basic variations of supergravity. This indicates the modification of the set of supergravity pre-potentials  and of the auxiliary field sector of minimal supergravity. The modified off--shell formulation,
the {\it special minimal supergravity}, can be obtained from the generic minimal supergravity by replacing one of its scalar auxiliary fields by a divergence of an auxiliary vector \cite{Ovrut:1997ur}.
The change in prepotential structure which characterizes the special minimal supergravity is that its  chiral compensator superfield is a {\it special chiral superfield}; this is to say it is constructed from a real prepotential superfield  rather than from the complex prepotential as it is the case for the chiral compensator of generic minimal supergravity  \cite{Kuzenko+05}.

To make the above statements clearer, let us turn to their flat superspace counterparts.

The differential forms in flat $D=4$, ${\cal N}=1$ superspace  had been studied already in early 80s \cite{Gates:1980ay,Gates:1980az}  (see also \cite{Binetruy:1996xw,Ovrut:1997ur}). In particular, it was known that  one can construct the closed 4--form $H_4(\Phi)$ in term of generic chiral superfield $\Phi$
(obeying $\bar{D}_{\dot{\alpha}}\Phi=0$) and its c.c. $\bar{\Phi}$  (obeying ${D}_{{\alpha}}\Phi=0$; here and below ${D}_{{\alpha}}$ and $\bar{D}_{\dot{\alpha}}$ are flat superspace covariant derivatives which obey $\{ {D}_\alpha ,\bar{D}_{\dot\alpha}\} = 2i \sigma^a_{\alpha {\dot\alpha}}{\partial}_a $). The general solution of the chirality conditions $\bar{D}_{\dot{\alpha}}\Phi=0$ is given by $\Phi = \bar{D}\bar{D}{\cal U}\equiv  \bar{D}_{\dot{\alpha}}\bar{D}{}^{\dot{\alpha}}{\cal U}$ where ${\cal U}$ is an unconstrained complex superfield called pre-potential. The expression for the corresponding three from potential $C_3(\Phi)$ such that $H_4(\Phi)=dC_3(\Phi)$ can be also found in
\cite{Gates:1980ay,Gates:1980az},  but this is given in term of real superfield $V=V^*$  and are valid if this real superfield serves as a pre-potential for the chiral superfield $\Phi^\prime$ in terms of which the closed four form $H_4(\Phi^\prime)$ is constructed, $\Phi^\prime = \bar{D}\bar{D}V$,
$\bar{\Phi}^\prime =DD V$. In this sense one can state that the superspace three form gauge theory is dual to a {\it special chiral multiplet}, constructed from the real pre-potential, rather than to a generic scalar supermultiplet described by the generic chiral superfield.

This fact was not stressed in \cite{Gates:1980ay,Gates:1980az}, probably because the field content of special chiral superfield $\Phi^\prime = \bar{D}\bar{D}V$, is very close to the generic scalar supermultiplet (described by $\Phi = \bar{D}\bar{D}{\cal U}$ with complex  ${\cal U}$): all the difference is in that one of the auxiliary scalar fields is replaced by the divergence of the auxiliary vector field, $\partial_\mu k^\mu$.
Such a difference becomes essential when the Lagrangian description of the supersymmetric matter interacting with supermembrane is studied \cite{IB+CM=2010}: when the supermembrane Wess--Zumino term is constructed with the use of real prepotential $V$, the (super)field theoretical part of the interacting action should include special chiral multiplet described by the special chiral superfield $\Phi^\prime = \bar{D}\bar{D}V$ rather than generic chiral multiplet.

Similarly, as far as the action of supermembrane interacting with supergravity, Eq. (\ref{Sint=SG+Sp2}), is constructed with the use of three form potential $C_3$ characterized by Eqs. (\ref{H4=HL+HR}), (\ref{vC3:=}), (\ref{bff*f=}), (\ref{bffb=}), (\ref{bfbb=}), the (super)field theoretical part of the supergravity--supermembrane interacting system should contain the special minimal supergravity, described by the action (\ref{SGact}) and admissible supervielbein variations (\ref{varEa}), (\ref{varEal}) and  (\ref{DDcU=vV+})  (rather than by generic minimal supergravity for which the restriction (\ref{DDcU=vV+}) is not present and the complex variation $\delta {\cal U }$ remains arbitrary).
The detailed study of the superfield equations of these dynamical system will be the subject of forthcoming paper \cite{IB+CM=2011}. Here we have reviewed the special minimal supergravity of \cite{Ovrut:1997ur}, describing the corresponding admissible variations of the supervielbein and three form potential $C_3$, and, using these, have derived the explicit form of the supermembrane supercurrent  and supergravity superfield equations of motion with supermembrane contribution.

Using Eqs. (\ref{varEa}), (\ref{varEal}) and  (\ref{DDcU=vV+}) we shown that the superfield equations of motion of the special minimal supergravity, allowing for the dynamical coupling to supermembrane, are modified, with respect to the ones of the generic minimal supergravity, by the presence of an arbitrary real constant $c$ (appearing as integration constant). This in its turn results in modification of the Einstein equation of supergravity by inclusion of a cosmological constant term proportional to $c^2$, so that in special minimal supergravity the cosmological constant is generated dynamically. In the language of component approach the dynamical generation of cosmological constant in special minimal supergravity was found already in \cite{Ovrut:1997ur}. In the superfield perspective such a mechanism of the cosmological constant generation was found for the first time by Ogievetsky and Sokatchev \cite{OS80} in the frame of their formulation of supergravity as the  theory of axial vector superfield \cite{OS78}.

The present contribution provides a basis for the study of superfield Lagrangian description of the more general $D=4$ interacting system including  supermembrane, supergravity and matter multiplets (see \cite{Tomas+} for the spacetime component study), in particular of the $D=4$ representative of the family of ''curious supergravity theory'' described in \cite{Duff+Ferrara=2010}.

\bigskip

{\it Acknowledgments}. The authors are thankful to Sergey Kuzenko, Paul Townsend  for useful discussion and communications, and to Daniel Waldram for bringing to our attention very important references \cite{Ovrut:1997ur} and \cite{Binetruy:1996xw}.  The partial support by the research grants FIS2008-1980 from the Spanish MICINN  and by the Basque Government Research Group Grant ITT559-10 is greatly acknowledged.

 \end{document}